\documentclass{jfm_mod}

\usepackage{graphicx}
\usepackage{natbib}
\usepackage{amsmath,amssymb}
\usepackage{color}
\usepackage{changes}
\usepackage{pdflscape}

\newcommand{\be}{\begin{equation}}
\newcommand{\ee}{\end{equation}}
\newcommand{\bs}{\begin{subequations}}
\newcommand{\es}{\end{subequations}}

\newcommand{\bd}[1]{{\boldsymbol #1}}
\newcommand{\bk}{\bd{k}}
\newcommand{\br}{\bd{r}}

\newcommand{\hzeta}{\hat{\zeta}}
\newcommand{\hp}{\hat{p}}
\newcommand{\hw}{\hat{w}}
\newcommand{\hpext}{\hp_\text{ext}}

\newcommand{\rmd}{\mathrm{d}}
\newcommand{\rme}{\mathrm{e}}
\newcommand{\rmi}{\mathrm{i}}

\newcommand{\pext}{p_\text{ext}}
\newcommand{\kint}{\int\frac{\rmd^2 k}{(2\pi)^2}}

\definecolor{burgundy}{rgb}{0.5, 0.0, 0.13}

\newcommand{\rev}[1]{{\color{burgundy} #1}}

\ifCUPmtlplainloaded \else
  \checkfont{eurm10}
  \iffontfound
    \IfFileExists{upmath.sty}
      {\typeout{^^JFound AMS Euler Roman fonts on the system,
                   using the 'upmath' package.^^J}%
       \usepackage{upmath}}
      {\typeout{^^JFound AMS Euler Roman fonts on the system, but you
                   dont seem to have the}%
       \typeout{'upmath' package installed. JFM.cls can take advantage
                 of these fonts,^^Jif you use 'upmath' package.^^J}%
       \providecommand\upi{\upi}%
      }
  \else
    \providecommand\upi{\upi}%
  \fi
\fi

\ifCUPmtlplainloaded \else
  \checkfont{msam10}
  \iffontfound
    \IfFileExists{amssymb.sty}
      {\typeout{^^JFound AMS Symbol fonts on the system, using the
                'amssymb' package.^^J}%
       \usepackage{amssymb}%
         \let\leq=\leqslant
         
      }{}
  \fi
\fi

\ifCUPmtlplainloaded \else
  \IfFileExists{amsbsy.sty}
    {\typeout{^^JFound the 'amsbsy' package on the system, using it.^^J}%
     \usepackage{amsbsy}}
    {\providecommand\boldsymbol[1]{\mbox{\boldmath $##1$}}}
\fi

\title[Observation of surface wave patterns modified by sub-surface shear currents]{Observation of surface wave patterns modified by sub-surface shear currents}
\author[B. K. Smeltzer, E. \AE s\o y \& S. \AA. Ellingsen]{Benjamin K.~Smeltzer$^1$, Eirik {\AE}s{\o}y$^1$, Simen \AA. Ellingsen$^1$\thanks{Email address for correspondence: simen.a.ellingsen@ntnu.no}}

\affiliation{$^1$Department of Energy and Process Engineering, Norwegian University of Science and Technology, N-7491 Trondheim, Norway}

\begin{document}
\maketitle

\begin{abstract}
  We report experimental observations of two canonical surface wave patterns --- ship waves and ring waves --- skewed by sub-surface shear, thus confirming effects predicted by recent theory. 
  Observed ring waves on a still surface with sub-surface shear current are strikingly asymmetric, an effect of strongly anisotropic wave dispersion. 
  Ship waves for motion across a sub--surface current on a still surface exhibit striking asymmetry about the ship's line of motion, and large differences in wake angle and transverse wavelength for upstream vs downstream motion are demonstrated, all of which in good agreement with theoretical predictions.  
  Neither of these phenomena can occur on a depth-uniform current. 
  A quantitative comparison of measured vs predicted average phase shift for a ring wave is grossly mispredicted by no-shear theory, but in good agreement with predictions for the measured shear current. A clear difference in wave frequency within the ring wave packet is observed in the upstream vs downstream direction for all shear flows, while it conforms with theory for quiescent water for propagation normal to the shear current, as expected. Peak values of the measured 2-dimensional Fourier spectrum for ship waves are shown to agree well with the predicted criterion of stationary ship waves, with the exception of some cases where results are imperfect due to the limited wave-number resolution, transient effects and/or experimental noise. 
  Experiments were performed on controlled shear currents created in two different ways, with a curved mesh, and beneath a blocked stagnant-surface flow. Velocity profiles were measured with particle image velocimetry, and surface waves with a synthetic schlieren method.   Our observations lend strong empirical support to recent predictions that wave forces on vessels and structures can be greatly affected by shear in estuarine and tidal waters. 
\end{abstract}

\section{Introduction}

Ship waves and ring waves have been admired since the dawn of time, and 
studied scientifically for 
two centuries \citep{wehausen60}, pioneered by \citet{cauchy1827}, \citet{poisson1816} and Kelvin \citep{kelvin1887}.
Observations of their modification by sub-surface shear flows, however, have not been reported, despite the abundance of flows with strong vertical shear in nature, for instance beneath the wind-swept ocean surface and in coastal and riverine waters. 
In these waters understanding the physics of surface waves can be crucial to a wide range of problems, including wave loads on vessels and installations, transportation of nutrients and pollutants, and thermal mixing, a key factor in climate modelling \citep{belcher12}. 
It has long been recognized that surface waves are profoundly affected by sub-surface shear \citep{peregrine76}, yet only recently has sub-surface shear been implemented in widely used ocean models \citep{kumar11,elias12}, and remote sensing of near-surface shear currents has been a point of attention in the last few years \citep{lund15,campana16}.

Recent theory has predicted that ship and ring waves are affected in striking ways by sub-surface shear; for instance, ring waves such as from a pebble thrown into the water no longer form concentric rings \citep{johnson90,ellingsen14b, khusnutdinova16}, and Kelvin's famous result that ship waves in deep water always form an angle of approximately $39^\circ$ no longer holds \citep{ellingsen14a, li16}. 
The physical understanding of ship waves more generally has been the focus of much recent attention \citep{rabaud13, darmon14,noblesse14,pethiagoda14}.

In the strongly sheared waters of the Columbia River delta 
--- often called ``the Graveyard of the Pacific'' for its dangerous sailing conditions and many shipwrecks --- 
the steepness of breaking waves 
was recently found by \cite{zippel17} to be mispredicted by up to $20\%$ unless shear was accounted for. In the same waters the wave-making resistance on smaller ships, typically accounting for more then $30\%$ of fuel consumption \citep{faltinsen05}, was found to differ by a factor $3$ or more between upstream and downstream motion at the same typical operational velocity relative to the water surface \citep{li17}. Studying wave forces during hurricanes theoretically 45 years ago, \citet{dalrymple73} found large shear effects and concluded that ``It is obvious ... that rational offshore design must include the effect of [shear] currents''. This, however, is still not the practice today.

In this 
article 
we report observations of ring waves and ship waves skewed by shear --- the first to our knowledge --- and show that these are well described by recently developed theory. In the following we 
briefly review the theory for calculating linear wave patterns on shear currents of arbitrary profiles, for comparison with our measurements. We next 
describe the experimental set--up, whereupon observations of ring waves and ship waves are presented and discussed, in that order. We conclude with a discussion of implications of the results in a wider context.

\section{Theoretical prediction of wave patterns}
\label{sec:theory}

We will briefly review the theoretical framework for the calculation of transient, general linear wave patterns propagating upon a horizontal shear current of arbitrary velocity profile, formulated in a manner suited for the present purposes. The theory, suitable for ship waves in particular, was first developed for stationary ship waves by \citet{smeltzer17} and further generalized by \citet{li17}. It assumes the source of the waves --- in our case ship waves and ring waves --- is a patch of pressure, $\hpext(\br,t)$, of arbitrary shape in the horizontal plane $\br = (x,y)$, and arbitrary dependence on time $t$. The pressure is $P(\br,z,t) = -\rho g z + \hp(\br,z,t)$, the vertical displacement of the free surface is $\hzeta(\br,t)$, and the vertical velocity is $\hw(\br,z,t)$; $z$ is the vertical axis. We let the mean water depth be $h$, and the undisturbed surface be $z=0$.  Hatted quantities are presumed small, and governing equations and boundary conditions are linearized with respect to these; their corresponding quantities in Fourier $\bk$-plane are 
\be\label{fourier}
  [\hzeta,\hw,\hp](\br,z,t) = \kint [\zeta,w,p](\bk,z,t)\rme^{\rmi {\bk\cdot \br}}
\ee
(it is understood that $\hzeta$ and $\zeta$ are not functions of $z$). 
We denote the wave vector $\bk = (k_x,k_y)$ so that $k = |{\boldsymbol k}|$. We assume unidirectional ambient current $U(z)$ along the $x$-axis. Numerically, inverse Fourier transforms are calculated with the inverse fast-Fourier transform method (iFFT). In the following the dependence on $\bk$ of quantities in phase space is implied.

Eliminating horizontal velocity components from the linearized Euler equation according to the standard procedure \citep[c.f.\ e.g.\ ][]{shrira93} yields the boundary value problem 
\bs
\begin{align}
(U-c)(w''-k^2w) = \mathbf{U}''w;~~ &-h<z<0 , \label{rayleigh}
\\
(U-c)^2 w'- (g+\frac{\sigma k^2}{\rho})w- U'(U-c) w = 0;~~ &z=0, \label{BC}\\
w =  0; ~~&z=-h,
\end{align}
\es
where it is understood that the perturbation quantities have time dependence $\exp[-\rmi \omega(\bk) t]$, and $c = \omega/k$ is the phase velocity. Using (\ref{rayleigh}) the boundary conditions \eqref{BC} can be combined into the implicit form \citep{ellingsen17}
\be
  (1+I)(c-U_0)^2 + (c-U_0)U_0' \tanh kh/k - c_0(k) = 0,
\ee
where 
\begin{align}
  I =& \int_{-h}^0 \rmd z \frac{U''(z)w(z,0)\sinh k(z+h)}{k(U-c)w(0,0)\cosh kh},
\end{align}
Here $c_0(k)=\sqrt{(g/k + \sigma k/\rho)\tanh kh}$, $U_0=U(0)$ and $U_0'=U'(0)$.

The resulting eigenvalues of $c$ are $c_\pm(\bk)$ and due to the relation $c_-(\bk) = -c_+(-\bk)$, only the positive value $c_+(\bk)$ is needed. There exists a number of different ways to calculate $c_+(\bk)$ numerically from an arbitrary $U(z)$, either numerically or using approximate analytical expressions --- for a review, see \citet{li18DIM} --- and how one chooses to calculate $c_+(\bk)$ is of no significance other than numerical cost and accuracy requirements. At the level of accuracy required for our present purposes an analytical approximation such as that of \citet{stewart74} for deep water would likely be adequate, yet introducing this additional source of error is unnecessary since we might instead employ, with little additional cost or complexity, the arbitrary-accuracy direct-integration method of \citet{li18DIM}. We hence assume $c_\pm(\bk)$ to be known functions henceforth.

The surface elevation has contributions from waves created at all previous times via a propagator function (Green function) $H_\zeta$, 
\be\label{zeta}
  \zeta(\bk,t) = \int_{-\infty}^t \rmd \tau \pext(\bk,\tau)H_\zeta(\bk,t-\tau),
\ee
hence all that is required is finding $H_\zeta$. It is shown by \citet{li17} that the solution may be written 
\newcommand{\tc}{\tilde{c}}
\be
  H_\zeta(\bk,t) = \frac{\tanh kh}{\rmi \rho(c_+-c_-)(1+I)}(\rme^{-\rmi kc_+t}-\rme^{-\rmi kc_-t})
\ee
with intrinsic phase velocity $\tc = c-U_0$, and $c_\pm$ are functions of $\bk$. Once $c_+(\bk)$ is known, $w(z)$ is found from \eqref{rayleigh} and $I$ is calculated. (An advantage of using the method of \citet{li18DIM} is that it builds on the same relations and yields $I$ together with $c_+$ automatically. )

If one assumes a stationary ship wave pattern, \eqref{zeta} can be simplified to 
\be\label{eq:thship}
  \zeta(\br) = \frac1\rho\lim_{\epsilon\to +0}\kint \frac{\pext(\bk)\tanh kh}{k(1+I)(c_+-\rmi\epsilon)(c_--\rmi\epsilon)}\mathrm{exp}(i \bk \cdot \br)
\ee
where a radiation condition has been applied. We use the fully transient expression however, to account for the linear acceleration phase of our model ship, since the transient wave pattern from this phase are visible in our experimental results for the two lowest Froude numbers we consider. 

Our ship wave predictions hypergaussian shape for the ship, 
\be\label{eq:pext}
  \hpext(\br) = p_0 \exp\left\{-\pi^2[(2x_\beta/L)^2+(2y_\beta/b)^2]^3\right\}
\ee
where the length and beam of the model ship are, respectively, $L=110$mm and $b=16$mm, and $x_\beta$, $y_\beta$ are coordinates relative to the ship with $x_\beta$ as the forward direction. 

For the initial value (ring wave) problem, a simpler form is more suitable. In fact, defining initial conditions for waves on shear currents in a fully consistent manner is a subtle affair, as discussed by \citet{akselsen18}, since some assumption must be made about the initial perturbation to the sub-surface vorticity field. However, we use the simplest choice, as used by \cite{ellingsen14b}\footnote{\citet{ellingsen14b} erroneously argued that this choice is necessary; see \citet{akselsen18}.} which makes for the following simple argument. We note that if $\hpext(\br,t)$ is supposed to be a short impulse at $t=t_0$, $\hp\propto \delta(t-t_0)$ the surface elevation \eqref{zeta} obtains the form (see \citet{li17})
\be\label{eq:prop}
  \zeta(\bk,t) = Z_+(\bk) \rme^{-\rmi kc_+(\bk)(t-t_0)} + Z_-(\bk) \rme^{-\rmi kc_-(\bk)(t-t_0)}
\ee
for $t>t_0$, with $Z_\pm$ as undetermined coefficients. It follows that any freely propagating wave pattern will be of this form once the source is ``switched off''. Assuming the surface elevation is known at $t=t_0$ to be $\hzeta(\br,0) = \zeta_0(\br)$ and $\partial\hzeta(\br,0)/\partial t = \dot{\zeta}_0$, \eqref{fourier} yields
\be\label{eq:icond}
  Z_\pm(\bk) = \mp\frac{ kc_\mp\zeta_0(\bk) - \rmi\dot{\zeta}_0(\bk)}{k[c_+(\bk) - c_-(\bk)]}.
\ee

\subsection{Initial conditions for the prediction of ring waves}
\label{sec:gs}

%

The initial conditions used for comparison with experiment are the values of $\zeta_0$ and $\dot{\zeta}_0$ at the earliest time where measurements are available. The surface elevation is found from the measured surface gradient field, sampled at discrete points in time (detail provided in \S \ref{sec:lab}), and $\dot{\zeta}_0$ may be approximated using a center difference scheme from the values of $\zeta$ at adjacent sampled points in time, accurate to second order in the small parameter $f_s T$ where $f_s$ is the camera sample rate and $T$ a characteristic timescale for the ring wave development. The approximation of $\dot{\zeta}_0$ is further improved using an algorithm similar to Gerchberg--Saxton (GS) \citep[c.f., e.g.,][]{fienup82}. The center difference approximation to $\dot{\zeta}_0$ is taken as an initial guess, and used with $\zeta_0$ in (\ref{eq:prop}) to propagate forward a value $t_s = f_s^{-1}$ in time to the next sampled surface elevation. The resulting calculated surface elevation is replaced by the measured values in (\ref{eq:icond}), whereby (\ref{eq:prop}) is used to propagate backwards a time $t_s$. The measured surface elevation is again substituted into (\ref{eq:icond}), and the entire process repeated multiple times. Since the surface elevation at subsequent times is determined by $\zeta_0$ and $\dot{\zeta}_0$, the method effectively finds the unknown $\dot{\zeta}_0$ from measurements of the surface elevation at two points in time.

\section{Laboratory set--up} 
\label{sec:lab}

\begin{figure}
  \centering \includegraphics[width=\textwidth]{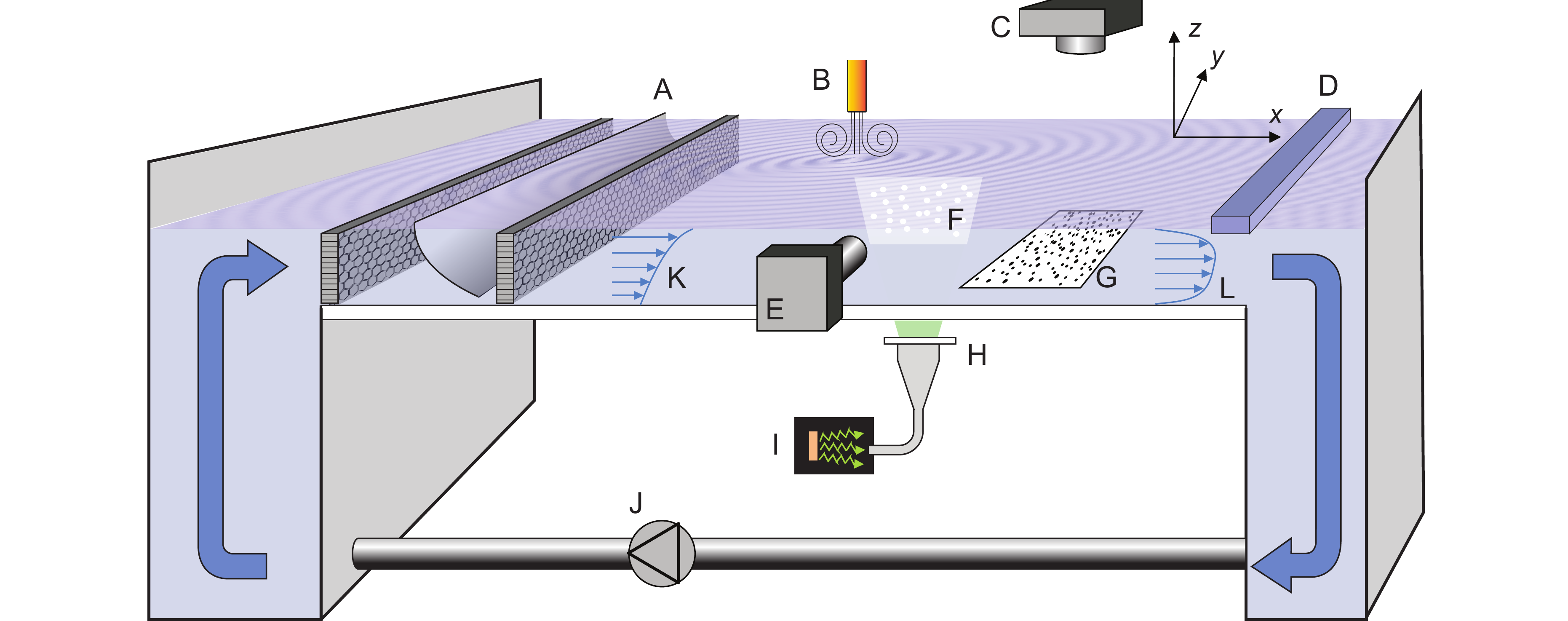}
  \caption{Laboratory set--up. A: curved mesh and honeycomb flow straighteners; B: pneumatic wave maker; C: Schlieren camera; D: stagnation bar; E: PIV camera; F: PIV light sheet; G: random dot pattern; H: collimation lens; I: LED light source; J: pump; K: shear profile due to curved mesh; L: stagnation shear profile.
}
  \label{fig:lab}
\end{figure}

The laboratory set--up is shown schematically in figure~\ref{fig:lab}. 
Shear currents were generated in a laboratory wave tank using a pump system to induce flow over a transparent plate of dimension $2$ m$\times 2$ m. 
The vertical profile of the streamwise current was measured using a 
particle image velocimetry (PIV) system 
similar to that described in \cite{willert10}, consisting of a high power LED light source (Luminus PT-120-TE),  fiber array, and collimation lens to produce a pulsed light sheet.  
This PIV system was able to measure the velocity profile at any location in a measurement area of approximately $50\text{cm}\times 50\text{cm}$.
The water was seeded with 40 $\mu$m polystyrene spheres (Microbeads AS) and the streamwise fluid velocity component was obtained by processing images from a camera (Imperx Bobcat 1610) mounted out of the plane as shown in figure~\ref{fig:lab}. The position of the light sheet can be moved to any location in the measurement area. 
The free surface elevation was measured using a synthetic schlieren (SS) method similar to that of \cite{moisy09}, imaging a random dot pattern mounted below the transparent plate (optical path length $\approx 2$ m) at a framerate of 35 Hz. 
Images were processed using a windowed digital image correlation technique, where the displacement of the dots relative to a still--water reference determines the local free--surface gradient, with horizontal spatial resolution of 5.7 mm. 
The free-surface elevation was found as the solution to an overdetermined linear system formed by expressing the gradients at each measurement point as second-order center differences in terms of the unknown surface elevation values\citep{moisy09}. 
The value of the surface elevation at a corner of the spatial domain was set to zero (effectively a necessary integration constant). As a result, there is a time-dependent offset in the mean surface elevation when waves pass over the corner point. The time-dependent offset is later removed by filtering, as described below.

The uncertainty in the measured gradients was estimated to be 0.001 or less in magnitude, based on analysis of a sequence of images taken with a still water surface. To estimate the corresponding error in the reconstructed surface elevation, an empirical law proposed by \citep{moisy09} was used. Assuming a root-mean-square (RMS) wave slope of 0.02 (a typical value, see e.g. figure~\ref{fig:ivp_line}a-d) and thus a relative uncertainty of 5\% in the measured gradients, the fractional error in the surface elevation was estimated to be 5\% for a 2.5 cm wavelength, and 1\% for a 10 cm wavelength. It is noted than in isolated regions of the measurement area containing waves with high curvature (particularly short wavelengths), larger errors may result due to lens effects of the dot pattern \citep{moisy09}.

To remove the offset in mean surface elevation as well as other noise, the data was filtered to remove periodic components that do not approximately satisfy the linear dispersion relation $\omega_0(\bk)$ in quiescent waters, allowing for deviations as expected due to the shear currents. The filter $F$ was defined in wavevector-frequency Fourier space of the surface elevation as
\be\label{eq:drfilt}
F(\bk,\omega) = \mathrm{exp}\left[-\left(\frac{\omega-\omega_0(\bk)}{U_{\mathrm{max}}k}\right)^4\right],
\ee
where $\omega$ is the frequency, $k = |\bk|$, and $U_{\mathrm{max}}$ is a maximum expected current. The surface elevation and gradient fields were filtered by taking the Fourier transform in spatial dimensions and time, multiplying the result by (\ref{eq:drfilt}), and performing an inverse Fourier transform of the product back to the spatial-temporal domain. The filter removes periodic components that lie outside a frequency range $U_{\mathrm{max}}k$ from the quiescent water dispersion relation. The value of $U_{\mathrm{max}}$ was chosen in practice to be $\approx 0.05$ $\mathrm{ms}^{-1}$ larger than the maximum current measured by PIV.

\begin{figure}
  \centering \includegraphics[width=\textwidth]{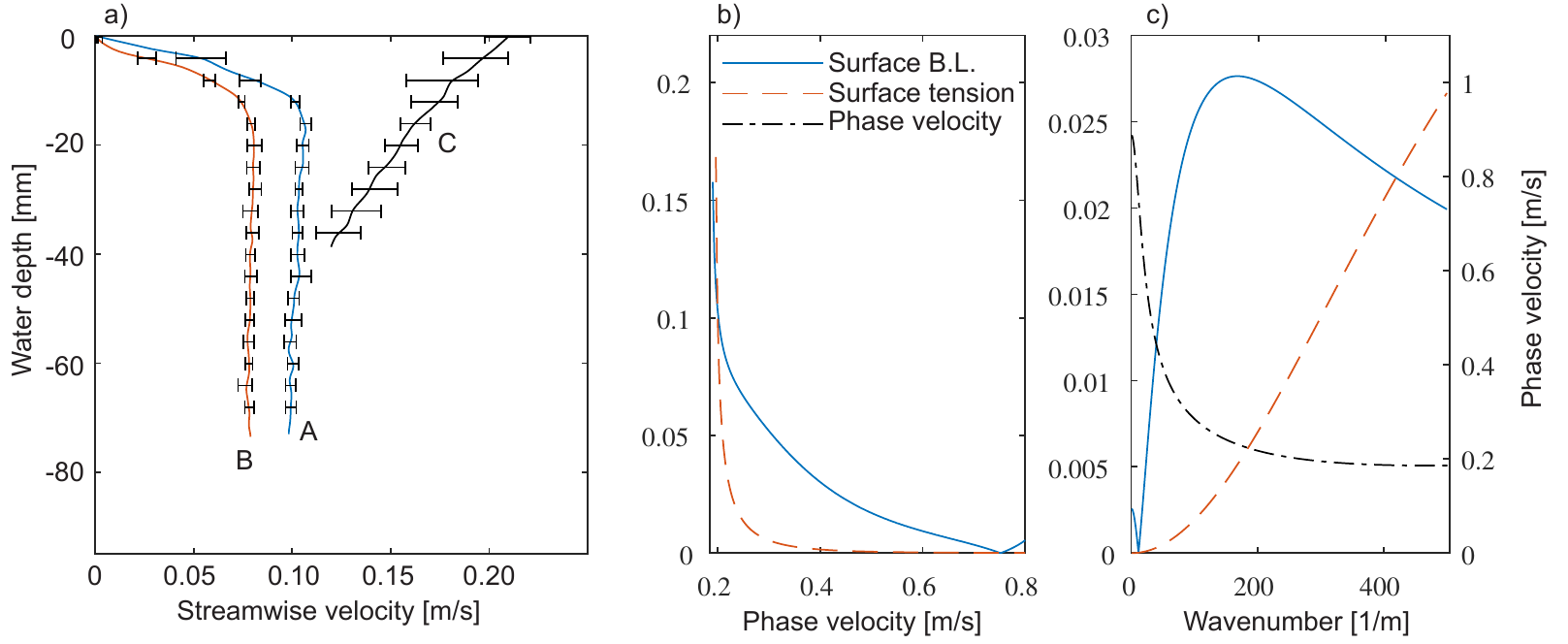}
  \caption{a):   
  Measured velocity field in stagnation region (profiles A and B) and downstream of curved mesh (profile C). b)-c) The alterations in wave dispersion within the stagnation region due to the measured surface boundary layer evolution as well as the predicted variation of the surface tension over the measurement area. The fractional variation in wavenumber as a function of phase velocity across the streamwise boundaries of the measurement region is shown in b), while the fractional variation in phase velocity as a function of wavenumber is shown in c), along with the phase velocity (dashed-dotted curve) for which the rightmost vertical axis applies. Downstream (shear-assisted) propagating waves are considered. The legend in b) applies to b)-c).   
}
  \label{fig:pivih}
\end{figure}

Shear current profiles were created 
in two ways, each 
producing conditions with good 
streamwise and spanwise homogeneity
within the measurement area. Firstly, 
similar to the method of e.g.\ \citet{dunn07}, 
a curved mesh at the inlet was used to create a moderately sheared 
current of approximately linear profile 
with vorticity vector along the positive $y$-axis,
as shown in figure~\ref{fig:lab}b. 
Secondly, 
an area exists covering roughly the downstream half of the tank, where
the 
surface is stagnant in the lab frame of reference, with the flow dipping suddenly beneath it developing a surface boundary layer.
This ``stagnation region'' stretches across the spanwise range and reaches up to a sharp spanwise separation line --- a Reynolds ridge \citep{scott82} --- some $80$ to $200$ cm upstream, depending on flow rate and the concentration of surface contaminants.  
The area can be moved further upstream with the simple placement of a horizontal spanwise bar dipping slightly beneath the surface. 

The velocity profiles used herein, measured with PIV, are shown in figure~\ref{fig:pivih}a for stagnation region profiles (A and B) and the flow due to the curved mesh (profile C). Error bars shows maximum and minimum values within the measurement area measured at four points, one in each quadrant, with the graph showing the average. The streamwise variation of the shear profiles A and B is in rough agreement with the expected boundary layer thickness as a function of distance downstream from the Reynolds ridge. The depth was 
$95$mm for stagnation region flow, and $50$mm for curved mesh flow; velocity profiles were not measured for the bottom $15$mm or so, yet the relevant wave spectra have hardly any contribution from waves long enough to be influenced by the bottom boundary layer.

Ring waves were produced using a pneumatic wave maker discharging bursts of air onto the surface, causing negligible disruption of the ambient current. The air flow nozzle has diameter $1.0$ mm and was positioned $50$-$100$ mm above the water surface, and the air flow was controlled by a solenoidal valve in $50$ ms pulses. 
For the curved mesh flow, profile C of figure~\ref{fig:pivih}a, the nozzle was moved along with the flow at the surface velocity using a linear stage. 
Ship waves were generated by mounting a miniature hull (Series 60) on a linear stage to move the ship at constant speed at different angles to the shear current. After initial acceleration of 1 $\mathrm{ms}^{-2}$, velocity variations were found to be less than $1$\% of the design speed.

The surface tension coefficient was determined by measuring waves produced by pulsing the pneumatic wavemaker at frequencies in the range $5$-$10$ Hz in otherwise quiescent water, and analyzing the wavenumber-frequency spectrum obtained by performing a discrete fast Fourier transform in space and time, where peaks in the spectrum are expected to lie on the linear dispersion surface. A set of points $\{\omega_i, k_i\}$ were extracted from the data at wave frequencies $\omega_i$ by fitting a Gaussian function along various directions in $k_x$-$k_y$ space to find the peak $k_i$ of the spectral signal. The set of points was then fitted to the function $\omega(k) = \sqrt{(gk + \Gamma k^3)\tanh kh}$, with $\Gamma$ the free parameter representing the ratio of surface tension to fluid density.

In the stagnation region results the concentration of surface contaminants is higher than in regions where the surface is moving as well as in quiescent water when the pump is turned off, and consequently the surface tension may be notably affected. To obtain representative values of the surface tension with the pump turned on, horizontal bars spanning the width of the tank and dipping beneath the surface were placed at the upstream and downstream boundaries of the wave measurement region prior to turning off the pump. Thus the contaminants were effectively prevented from spreading over the entire channel region and a representative measurement of surface tension could be made on the now quiescent water in the same way as described in the previous paragraph. It is noted that within the surface stagnation region there is in fact a surface tension gradient in the streamwise direction, which arises to balance the surface shear stress from the shear current \citep{harper74}. Using a value of the viscosity in clean water and typical measured values of the surface shear, the surface tension coefficient was estimated to vary by approximately 0.003 $\mathrm{Nm}^{-1}$ across the wave measurement region, or $3\times 10^{-6}$ $\mathrm{m}^3\mathrm{s}^{-2}$ in the value of $\Gamma$ amounting to a relative variation of $\sim 10$\% over the measurement area. 

Both the variation of the surface tension as well as the shear flow within the measurement area contribute to an inhomogeneous wave dispersion relation with a slight $x$-dependence. The theoretical predictions used herein assume a dispersion relation that is homogeneous in space. Measures of the variation in the dispersion relation at the upstream and downstream boundaries of the wave measurement area based on the measured variations in shear profile B as well as the estimated variation in surface tension are shown in figure~\ref{fig:pivih}b-c. The fractional variation in wavenumber for a given phase velocity is shown in figure~\ref{fig:pivih}b, while the variation in phase velocity as a function of wavenumber is shown in figure~\ref{fig:pivih}c, where the leftmost vertical axis applies. Also shown in figure~\ref{fig:pivih}c is the phase velocity as a function of wavenumber (where the rightmost vertical axis applies) for reference, calculated assuming the average shear profile and surface tension coefficient over the measurement area. Downstream propagating waves are considered, though the trends are similar for upstream waves. For all but the shortest measured wavelengths, the surface boundary layer development has a greater effect on wave dispersion variation across the measurement domain. Given a minimum phase velocity of $\approx 0.19$ $\mathrm{ms}^{-1}$, the fractional error in wavenumber increases dramatically for phase velocities approaching the minimum value, where the phase velocity derivative with respect to wavenumber goes to zero. For measured ring waves, the surface boundary layer results in a $\leq2.5$\% variation in the phase evolution of a wave, resulting in wave components drifting slightly out of phase compared to theoretical predictions. In the context of ship waves measurements for relevant velocities of $0.3-0.6$ $\mathrm{ms}^{-1}$, there is a $\leq 5$\% variation in the wavelength of wave directly following ship. The variation in wave dispersion is expected to be one of the main causes of disagreement between experiment measurements and theoretical predictions shown in the following sections, and is discussed further in the context of the quantitative results shown in \S\ref{sec:ring} and \ref{sec:ship}. The assumption of a homogeneous dispersion relation in the theory is deemed satisfactory for the level of accuracy of the present study, especially considering that disagreements caused by the variations shown in figure~\ref{fig:pivih}b-c are partially averaged-out in quantitative metrics that analyze the waves over the entire measurement region.

The presence of increased surface contaminants in the stagnation region results in a thin viscoelastic surface film, which may notably increase the viscous damping of surface waves compared to water with a clean surface \citep{alpers89}. However, estimates of the damping from analysis of the wave measurements used in determining $\Gamma$ showed no clear trends of increased damping rates relative to theoretical predictions based on the Stokes equation \citep{alpers89} within the level of accuracy of the analysis. The inviscid theory presented in \S\ref{sec:theory} does not include damping effects (the phase velocities and wavevectors are assumed real quantities), which may contribute to a slight disagreement between experiments and theoretical predictions for the smallest wavelengths (a damping rate of $\sim 1$ $\mathrm{m}^{-1}$ was estimated for a 2 cm wavelength, and $\sim 0.2$ $\mathrm{m}^{-1}$ for a 4.5 cm wavelength using the lowest measured surface tension value). It is noted that measured damping rates could be included in the theory in \S\ref{sec:theory} as an imaginary component to $c_\pm(\bk)$ for use in potential future studies.

To avoid ambiguity we will use the convention of \citet{li17} and refer to shear--assisted (shear--inhibited) directions of motion, pointing along (against) the current direction in a reference system where the surface is at rest. In particular, the ``downstream'' direction is shear--assisted in the stagnation flow region (profiles A and B in figure~\ref{fig:pivih}a), but shear--inhibited in the region behind the curved mesh (profile C in figure~\ref{fig:pivih}a) as the current velocity decreases with depth relative to the value at the surface for profile C.

\newcommand{\Fr}{\mathrm{Fr}}

\section{Linear dispersion relation}

\begin{figure}
  \centering \includegraphics[width=\textwidth]{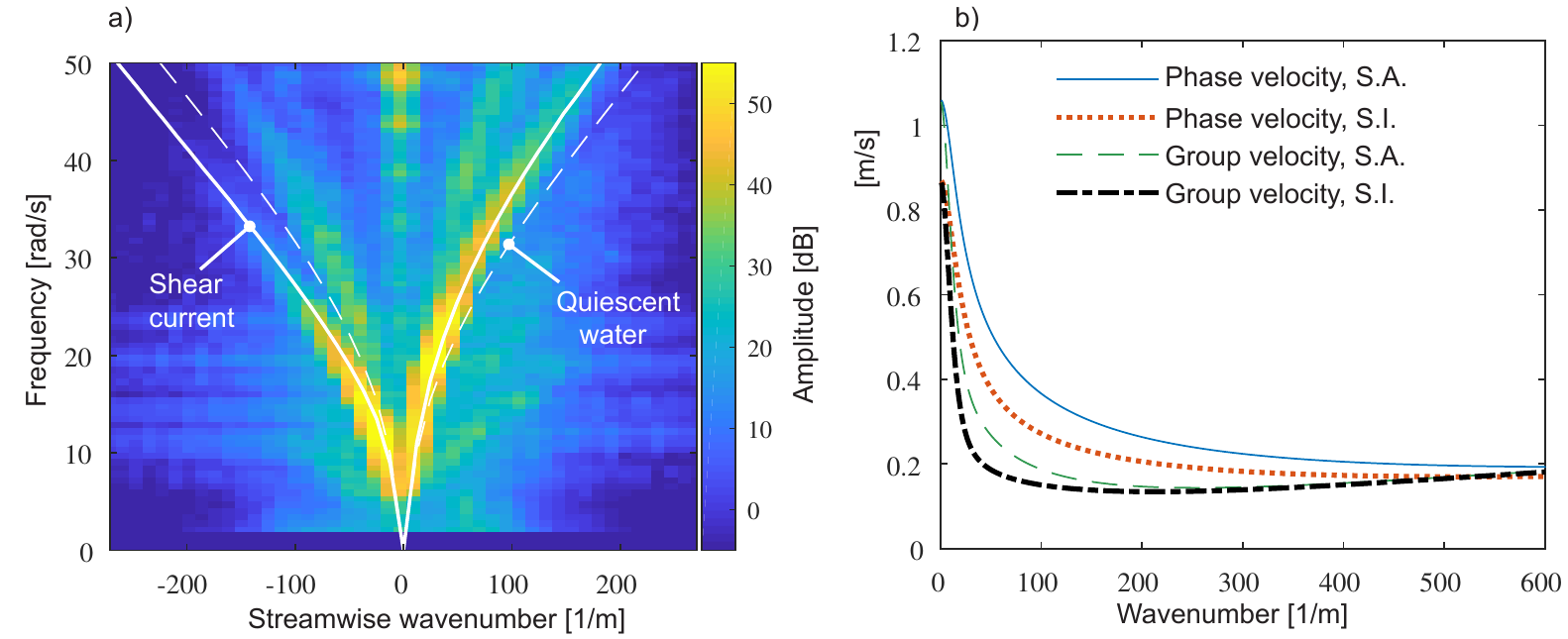}
  \caption{(colour online) a) Measured dispersion relation for the shear current in the stagnation region, profile A in figure~\ref{fig:pivih}a, and theoretical predictions for quiescent water and shear current, respectively. b) Calculated phase and group velocities for wave propagation in the shear-assisted (S.A.) and shear-inhibited (S.I.) directions atop the same shear profile.  }
  \label{fig:dr}
\end{figure}

We measured the dispersion relation for waves atop shear current profile A by driving omnidirectional waves approximately monochromatically over a range of frequencies, then Fourier transforming the surface gradients in time and space to obtain spectra $P_x(\bk,\omega)$ and $P_y(\bk,\omega)$ for the gradient components along the $x$ and $y$--directions respectively, where $\omega$ is the frequency. The magnitude squared of the spectra for each component is added together to produce a single spectrum shown in figure~\ref{fig:dr}a. 
The surface tension coefficient parameter was measured to be $\Gamma = 2.8 \pm 0.1 \times 10^{-5}$ $\mathrm{m}^3\mathrm{s}^{-2}$. A clear maximum of the power spectrum is seen along the dispersion curve
which is consistent with the dispersion relation $\omega(\bk)$ calculated numerically for the measured dispersion profile (solid line), but inconsistent with the quiescent water prediction (dashed line). Theoretical predictions of the phase and group velocities in the shear-assisted and shear-inhibited directions are shown in figure~\ref{fig:dr}b as a function of wavenumber. Due to the background shear current, there are distinct differences in the velocities depending on direction of wave propagation, indicating that wave patterns such as ring waves and ship waves explored in the following sections will be significantly modified.


\section{Ring wave observations}
\label{sec:ring}

\begin{figure}
  \centering \includegraphics[width=\textwidth]{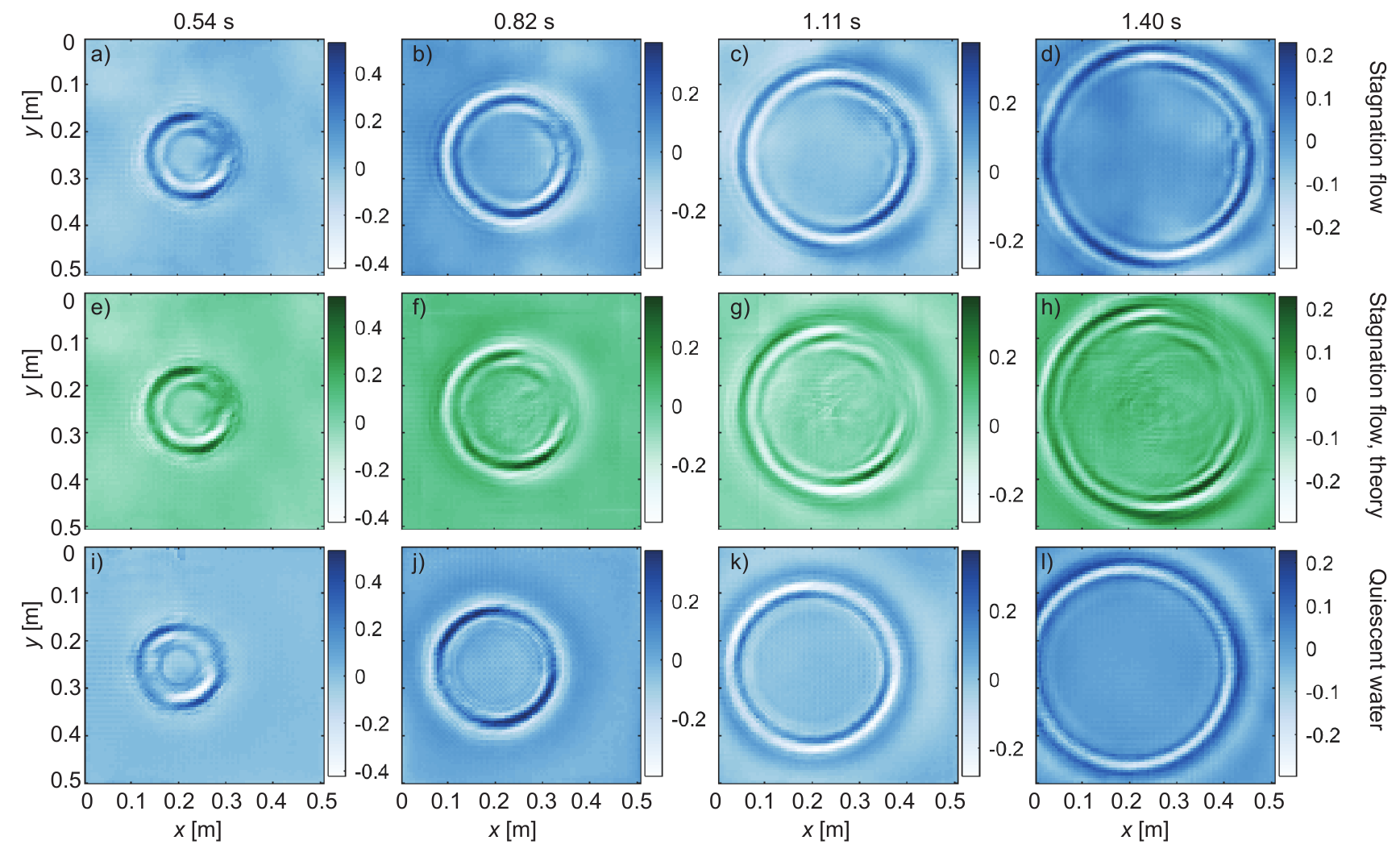}
  \caption{
	(colour online) 
  Ring waves, measured and predicted. Rows and columns respectively represent different cases and times after the initial pulse, as indicated. Blue-tinted panels are measured, green-tinted are theoretical predictions. The current is from left to right in the lab frame of reference. The colour-bars show the surface elevation in millimeters.
  }
  \label{fig:ringsr}
\end{figure}

\begin{figure}
  \centering \includegraphics[width=\textwidth]{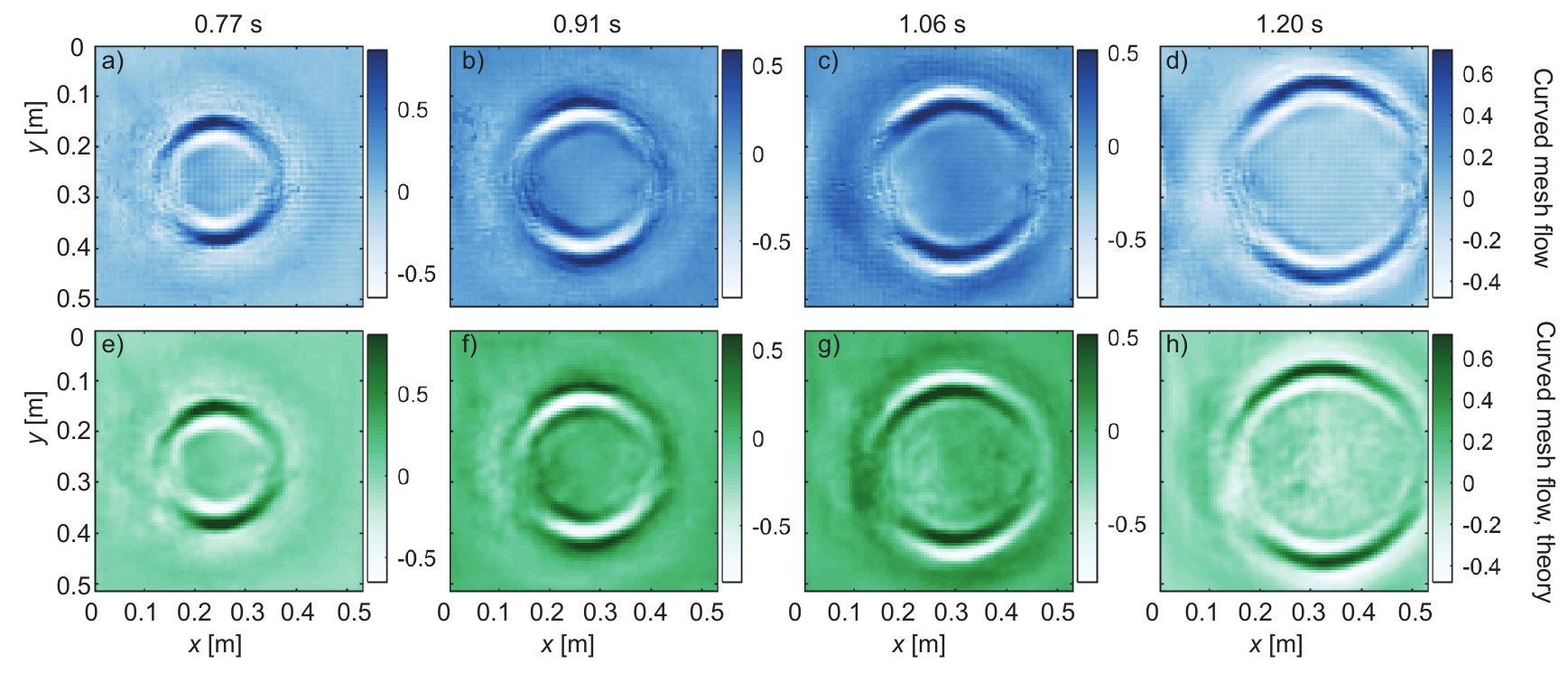}
  \caption{
	(colour online) 
  Ring waves, measured and predicted. Rows and columns respectively represent different cases and times after the initial pulse, as indicated. Blue-tinted panels are measured, green-tinted are theoretical predictions. The current is from left to right in the lab frame of reference. The colour-bars show the surface elevation in millimeters.
  }
  \label{fig:ringcm}
\end{figure}

Ring wave 
measurements atop the stagnation region flow profile A in figure~\ref{fig:pivih}a and corresponding theoretical predictions using (\ref{eq:prop}) and (\ref{eq:icond}) are shown in figure~\ref{fig:ringsr}. 
We compare measurements in quiescent water (figure~\ref{fig:ringsr}i-l with those made with the same initial pulse atop stagnation region shear profile A \rev{in figure~\ref{fig:pivih}a} (figure~\ref{fig:ringsr}a-d). Each column shows a particular time after the initial pulse, as indicated. The asymmetry 
compared to quiescent water is striking. 
Images of a ring wave propagating on the approximately linear shear current, profile C of figure~\ref{fig:pivih}a, are shown in figure~\ref{fig:ringcm}. The height of the pneumatic pressure pulse above the water surface was larger than for the stagnation region case, resulting in slightly longer wavelength components.
Due to the weaker shear the asymmetry is less conspicuous, yet differences between streamwise and spanwise propagation are still apparent. Left and right in the image 
appear to be 
reversed compared to the stagnation flow case since vorticity now has the opposite sign. 
The surface velocity is non-zero for the mesh flow, so the whole pattern is convected downstream (towards the right); this amounts to a change of reference system and can cause no asymmetry of the pattern itself. Further quantitative analysis presented below confirms the observable effects of the subsurface shear current. 
To ensure no asymmetry could arise due to relative motion of the surface and wave-maker, the pneumatic nozzle was moved along with the flow so as to follow the surface, using a linear stage.

\begin{figure}
  \centering \includegraphics[width=\textwidth]{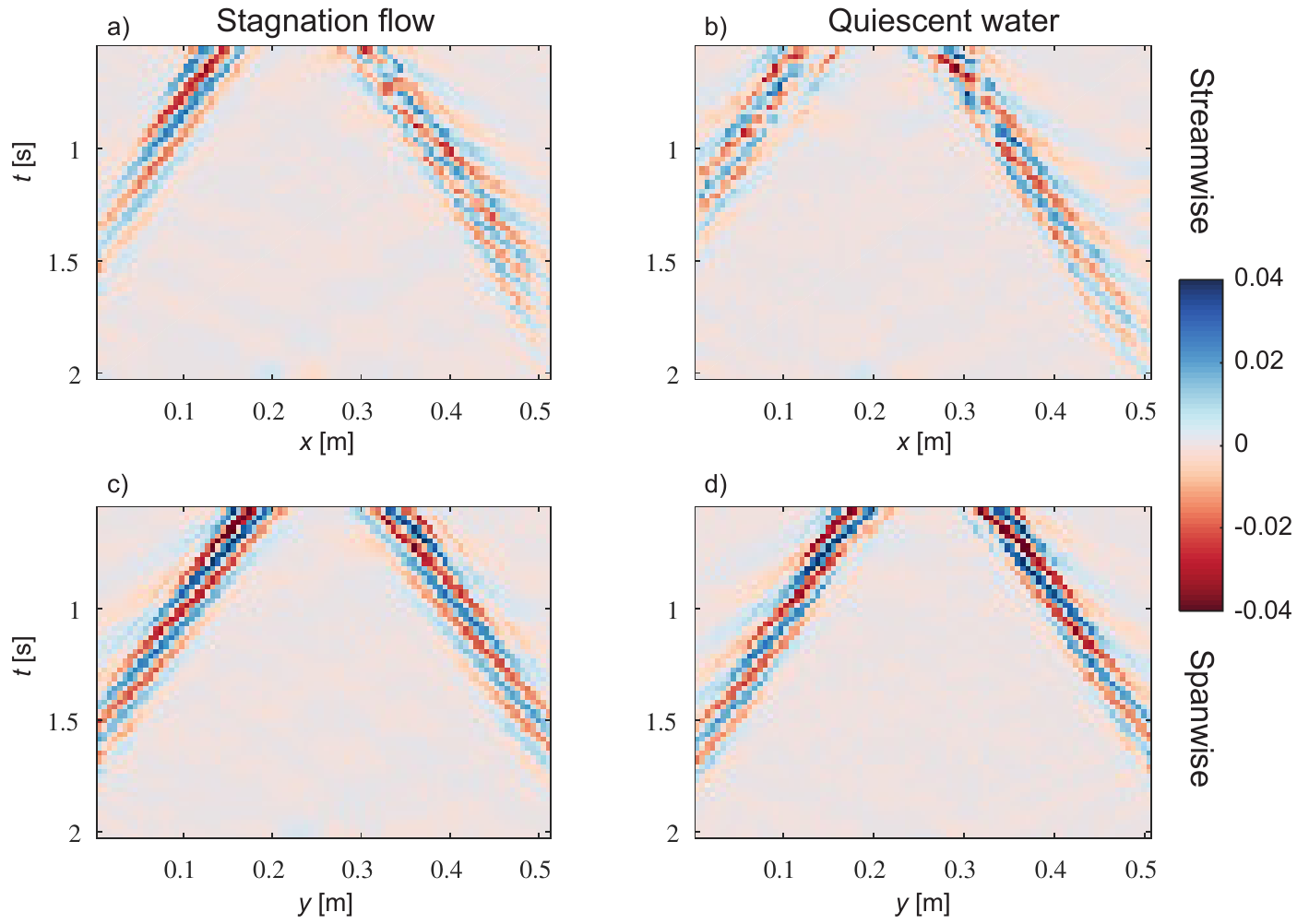}
  \caption{
	(colour online) 
  The measured surface gradient as a function of space and time for ring wave atop the subsurface shear current profile A in figure~\ref{fig:pivih}a ((a) and (c)), and in quiescent waters ((b) and (d)). (a) and (b) show the streamwise surface gradient component as a function of streamwise position $x$, along a line passing through the center of the initial pressure pulse. (c) and (d) show the same, but for spanwise positions and gradient component. The colour-bar applies to all panels.
  }
  \label{fig:ivp_line}
\end{figure}

As initial condition for the theoretical predictions, 
we use the measured surface elevation and velocity at 
the earliest available time, $t=0.54$ s for stagnation region flow, and $t=0.77$ s for the curved mesh profile. Before these times images are partly obscured by the pneumatic wave maker being within the field of vision. The coefficient representing the ratio of the surface tension coefficient to fluid density was found to be $\Gamma = 2.8 \pm 0.1 \times 10^{-5}$ $\mathrm{m}^3\mathrm{s}^{-2}$ for both the stagnation region and quiescent waters in figure~\ref{fig:ringsr}, reported to two significant figures. This corresponds to a surface tension coefficient of $0.028 \pm 0.001$ $\mathrm{Nm}^{-1}$ for a density of 997 $\mathrm{kg/m}^{3}$. The bounds indicate the 95\% confidence intervals of the fits. The stagnation region covered much of the length of the channel during the experiments, which is likely the reason the surface tension value is approximately unchanged when the pump was turned off. The parameter $\Gamma$ was not measured for the case of the curved mesh profile, and a value of $7\times 10^{-5}$ $\mathrm{m}^3\mathrm{s}^{-2}$ was used in the theoretical predictions, estimated based on other measurements of $\Gamma$ during the time period where the experiments were conducted.

The theoretically predicted surface amplitude at each respective time is 
shown in figure~\ref{fig:ringsr}e-h and \ref{fig:ringcm}e-h (green-tinted). 
Agreement with observations 
is qualitatively excellent, 
correctly predicting, in particular, the dispersion of the up-- and downstream propagating wave packets, both the distance traveled by the packet as a whole, the phase propagation of individual crests and troughs, and the dispersion of the packet. 

For a more detailed examination of the motion and width of the wave packets propagating outward from the initial pulse centered at spatial position $\br_0 = (x_0,y_0)$, we consider figure~\ref{fig:ivp_line} which shows the surface gradient components as a function along streamwise (figure~\ref{fig:ivp_line}a,b) and spanwise (figure~\ref{fig:ivp_line}c,d) lines passing through $\br_0$, showing positions upstream/downstream and orthogonally left and right of $\br_0$, respectively. Figures~\ref{fig:ivp_line}a and c are for the the ring waves measured atop the stagnation region flow, figures~\ref{fig:ivp_line}b and d for quiescent water (the same measurements shown in the top and bottom rows of figure~\ref{fig:ringsr}). Considering figure~\ref{fig:ivp_line}a, it is clear that the downstream propagating wave packet (shear-assisted direction) is greater in spatial width than the upstream packet (shear-inhibited direction), evidencing the increased (decreased) dispersion in the shear-assisted (inhibited) direction, confirming theoretical predictions of \citet{ellingsen14b}. For wave motion along the same direction in quiescent waters in figure~\ref{fig:ivp_line}b as well as wave motion in the spanwise directions in figure~\ref{fig:ivp_line}c-d, the spatial width of the wave packets in both directions is approximately the same. 

\begin{figure}
  \centering \includegraphics[width=\textwidth]{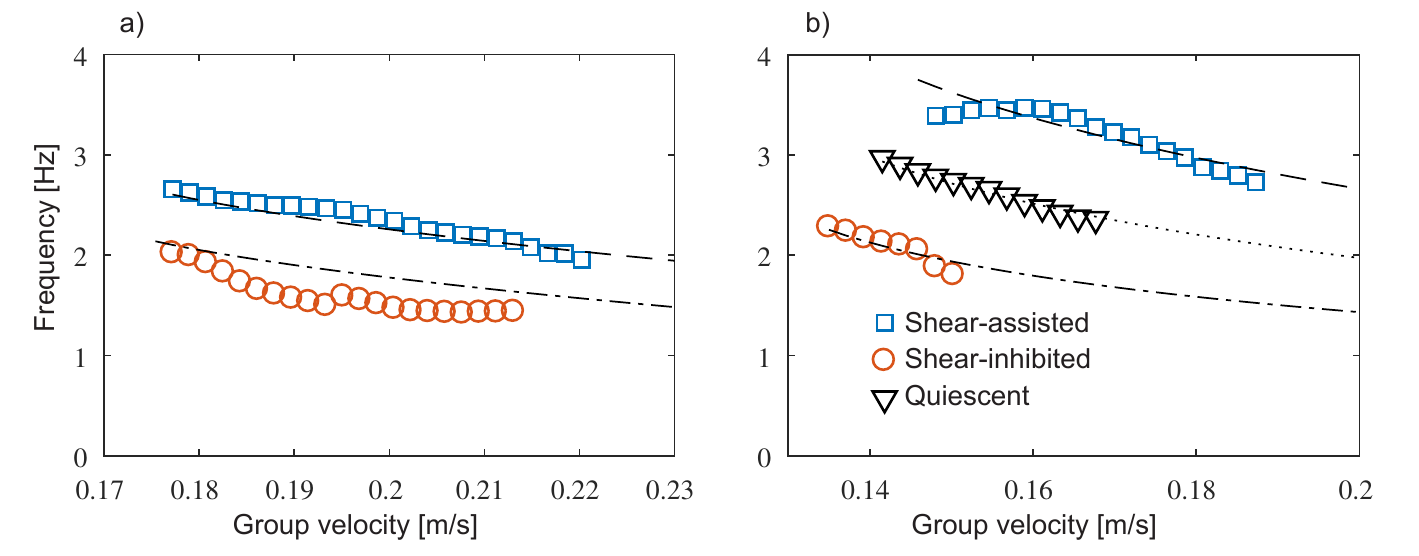}
  \caption{
	(colour online) 
  The frequency of waves traveling past an observer moving at a particular group velocity outward from the position of the initial ring wave pulse, for the curved mesh flow (a)) and stagnation region flow (b)). Shear-assisted and shear-inhibited direction are considered, as well as the case of quiescent waters for the stagnation region. The legend applies to both a) and b).
  }
  \label{fig:groupvel}
\end{figure}

The motion of the wave packet as a whole is determined by the group velocities of the relevant wavelengths, and is reflected by the slope of the packet as seen in figure~\ref{fig:ivp_line}, i.e. the translation of the packet in space over a given time interval. Also apparent in figure~\ref{fig:ivp_line} are darker and lighter lines along which the surface gradient has maximum, or zero, absolute value, respectively. These represent the motion of constant wave phase (for instance crests and troughs), and their slope indicates the phase velocities. Considering figure~\ref{fig:ivp_line}a, it is evident that while the group velocity is approximately the same in shear-assisted and shear-inhibited directions of propagation, the difference between phase and group velocity is greater (smaller) in the shear-assisted (shear-inhibited) direction. This confirms the theoretical predictions of \citep{ellingsen14b}. In figure~\ref{fig:ivp_line}b-d, the angular misalignment between lines representing the group and phase velocities is roughly the same in both propagation directions as expected. Further more quantitative analysis of the relationship between group and phase velocities for different propagation directions is given below in the discussion of figure~\ref{fig:groupvel}.

\begin{figure}
  \centering \includegraphics[width=\textwidth]{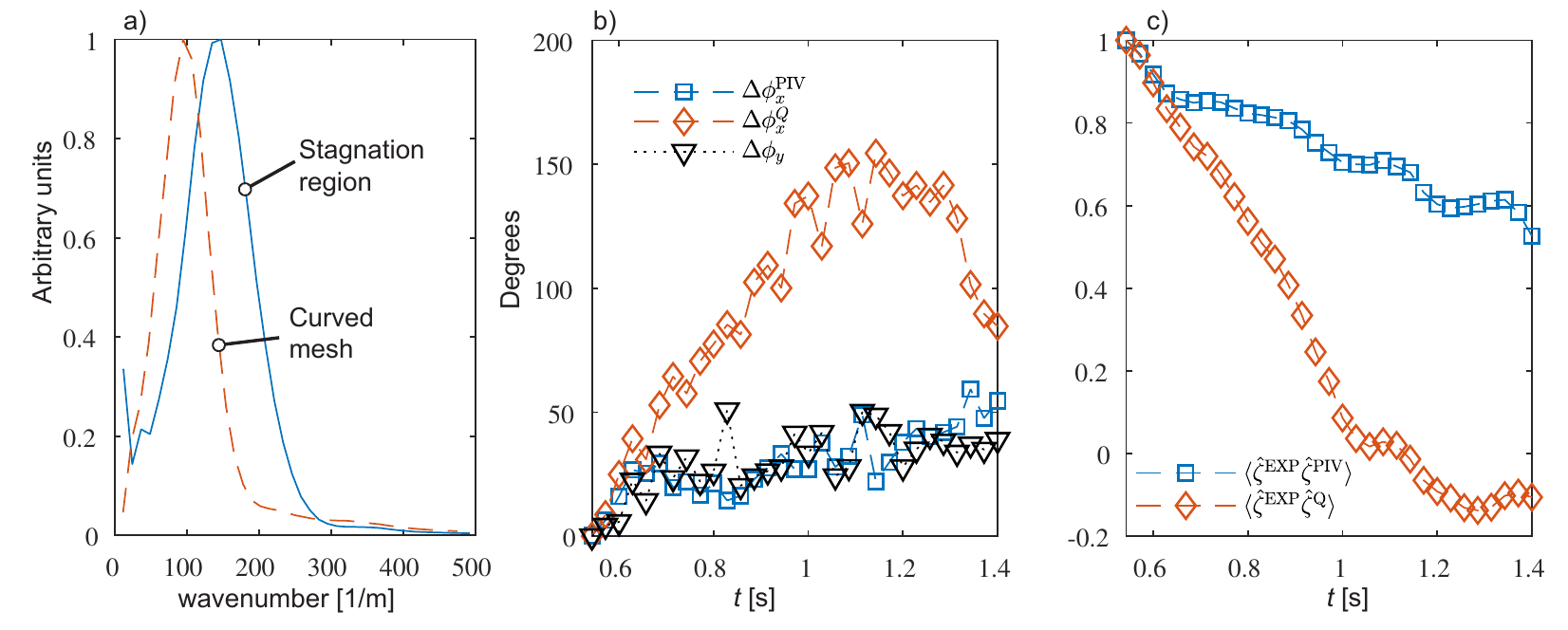}
  \caption{
	(colour online) 
  a): Wavenumber spectrum for ringwaves atop the curved mesh and stagnation region flows. b): Weighted average phase difference between ring waves measured experimentally in the stagnation region and theoretical predictions. $\Delta\phi_x^{\mathrm{PIV}}$ uses the dispersion relation for waves atop the shear profile as measured by PIV, considering waves along the $x$-direction. $\Delta\phi_x^{\mathrm{Q}}$ is the same but using the quiescent water dispersion relation. $\Delta\phi_y$ considers waves traveling in the $y$-direction perpendicular to the flow. c) Normalized inner product (defined in the text) between experimentally measured surface elevation, and that predicted by theory using the dispersion relation reflecting the measured shear profile ($\langle\hat{\zeta}^{\mathrm{EXP}}\hat{\zeta}^{\mathrm{PIV}}\rangle$) and quiescent dispersion relation ($\langle\hat{\zeta}^{\mathrm{EXP}}\hat{\zeta}^{\mathrm{Q}}\rangle$). For b) and c), initial time $t_0 = 0.54$ s.
  }
  \label{fig:ring2}
\end{figure}

We now proceed with some quantitative comparisons between the measured ring waves and theoretical predictions. For further use, we first define a wavenumber spectrum as:
\be\label{eq:kspec}
S(k) = \int_0^{2\pi}\int_0^\infty\mathrm{d}\omega\mathrm{d}\theta\left(|P_x(\bk,\omega)|^2 + |P_y(\bk,\omega)|^2\right),
\ee
with $\bk = k(\cos\theta, \sin\theta)$. The spectra $S(k)$ for the ring waves atop the stagnation region and curved mesh flow are shown in figure~\ref{fig:ring2}a, in units normalized to the peak value. As can be seen, the peak spectral components of the curved mesh flow are centered at a smaller wavenumber, consistent with the pneumatic pulse being positioned at a greater height, thus producing a broader initial deformation of the surface.

The theoretical predictions for ring waves use the measured surface elevation $\zeta_0$ and time derivative $\dot{\zeta}_0$ determined as in \S~\ref{sec:gs} as initial conditions in (\ref{eq:icond}), where $\zeta_0$ and $\dot{\zeta}_0$ are taken at the earliest time $t_0$ where the measured ring wave patterns were unobstructed by the pneumatic pulse mounting apparatus. The surface elevation at other times was predicted using (\ref{eq:prop}) and disregarding measurement errors in $\zeta_0$ and $\dot{\zeta}_0$, the accuracy of the theoretical predictions is related to the determination of phase velocities $c_\pm(\bk)$. Errors in $c_\pm(\bk)$ result in wavelength components drifting out of phase with time after $t_0$, reflected by the complex phase factors in (\ref{eq:prop}). To quantify this phase shift and assess the accuracy of the theoretical predictions, we define weighted averaged phase differences between the measured surface elevation for ring waves in the stagnation region, $\hat{\zeta}^{\mathrm{EXP}}$, and theoretical predictions with $c_\pm(\bk)$ calculated using the measured stagnation region shear profile ($\hat{\zeta}^{\mathrm{PIV}}$) and predictions using the quiescent water dispersion relation ($\hat{\zeta}^{\mathrm{Q}}$):

\bs
\begin{align}
\Delta\phi_x^{\mathrm{PIV}}(t) = \int_{k_{0.5}}\mathrm{d}kS(k)\mathrm{Arg}\left[ \zeta^{\mathrm{PIV}}(k,0,t)\zeta^{\mathrm{EXP},*}(k,0,t)\right]\bigg/\int_{k_{0.5}}\mathrm{d}kS(k),\label{dphix} \\
\Delta\phi_x^{\mathrm{Q}}(t) = \int_{k_{0.5}}\mathrm{d}kS(k)\mathrm{Arg}\left[ \zeta^{\mathrm{Q}}(k,0,t)\zeta^{\mathrm{EXP},*}(k,0,t)\right]\bigg/\int_{k_{0.5}}\mathrm{d}kS(k), \label{dphixq}\\
\Delta\phi_y(t) = \int_{k_{0.5}}\mathrm{d}kS(k)\mathrm{Arg}\left[ \zeta^{\mathrm{PIV}}(0,k,t)\zeta^{\mathrm{EXP},*}(0,k,t)\right]\bigg/\int_{k_{0.5}}\mathrm{d}kS(k),\label{dhpiy}
\end{align}
\es

where an un-hatted surface elevation quantity indicates a spatial Fourier transform as in (\ref{fourier}), an asterisk indicates a complex conjugate, $\mathrm{Arg}[]$ is the complex phase angle, and $k_{0.5}$ indicates integration over wavenumbers where $S(k)$ is greater than 50\% of the peak value. $\zeta^{\mathrm{EXP}}$, $\zeta^{\mathrm{Q}}$, and $\zeta^{\mathrm{PIV}}$ are functions of $(k_x,k_y,t)$. $\Delta\phi_x^{\mathrm{PIV}}(t)$ represents the spectrally-weighted phase difference between measured waves and theoretical predictions with $c_\pm(\bk)$ reflecting the measured phase velocity, for waves traveling along and against the streamwise direction ($k_y = 0$). $\Delta\phi_x^{\mathrm{Q}}(t)$ is the same but using the quiescent water dispersion relation in the theoretical predictions. $\Delta\phi_y(t)$ considers wave components propagating along the spanwise direction ($k_x = 0$), for which it is inconsequential which dispersion relation is used. The results for ring waves in the stagnation region are shown in figure~\ref{fig:ring2}b. $\Delta\phi_x^{\mathrm{Q}}$ initially increases at a much greater rate in time compared to $\Delta\phi_x^{\mathrm{PIV}}$ and $\Delta\phi_y$, since the dispersion relation doesn't include effects of the shear current. As the maximum phase difference defined here is $180^\circ$ which is a weighted average over different wavelength components drifting out of phase at varying rates in time, it is reasonable that the maximum of $\Delta\phi_x^{\mathrm{Q}}$ does not reach $180^\circ$. The decay of $\Delta\phi_x^{\mathrm{Q}}$ at larger times is likely due to wave components beginning to drift back in phase with the experimental measurements. The rate of increase of $\Delta\phi_x^{\mathrm{PIV}}$ and $\Delta\phi_y$ is roughly the same, likely caused by errors in the initial conditions used in (\ref{eq:icond}) and by inhomogeneity of the dispersion relation across the measurement area in the stagnation region as discussed in \S \ref{sec:lab}. Considering the peak wavenumber component in figure~\ref{fig:ring2}a, a 2.5\% variation in wave phase velocity (and frequency) is expected over the measurement area due to the surface boundary layer development (see figure~\ref{fig:pivih}c), which corresponds to a $\sim 50^\circ$ degree phase lag over the time period of figure~\ref{fig:ring2}b. This estimate is likely an upper-bound considering that a wave doesn't propagate over the entire measurement area, yet is roughly the peak value of the phase lag for $\Delta\phi_x^{\mathrm{PIV}}(t)$.

Another metric for quantifying the accuracy of theoretical predictions involves simply comparing the difference between predicted and measured surface elevation. We define the normalized inner product:

\be\label{eq:innerp}
\langle \hat{\zeta}^1 \hat{\zeta}^2  \rangle = \frac{\int \mathrm{d}\mathbf{r}\hat{\zeta}^1(\mathbf{r})\hat{\zeta}^2(\mathbf{r})}{ \sqrt{\int \mathrm{d}\mathbf{r}|\hat{\zeta}^1(\mathbf{r})|^2\int \mathrm{d}\mathbf{r}|\hat{\zeta}^2(\mathbf{r})|^2}},
\ee

where $\hat{\zeta}^1(\mathbf{r})$ and $\hat{\zeta}^2(\mathbf{r})$ are surface elevation functions of horizontal spatial coordinates at a particular time value. As for the phase differences, we consider theoretical predictions using dispersion relations with and without the effects of the subsurface shear current included. Inner products, between experimentally surface elevation and theoretical predictions as a function of time are shown in figure~\ref{fig:ring2}c. The inner product with the theoretical prediction assuming a quiescent water dispersion relation decays at a much higher rate than the case including the effects of shear currents, showing that the theoretical propagation in time of the initial perturbation is poor in the former case and far better in the latter. The necessity of taking shear-effects of the dispersion into account is again confirmed.

As predicted theoretically and observed qualitatively in figure~\ref{fig:ivp_line}, the difference between wave group and phase velocities is anisotropic in the presence of a subsurface shear current. To quantify the effect in the context of the experimentally measured ring waves, we consider an observer traveling at a particular velocity $V$ in a direction along the $x$-axis outward from the location of the initial pneumatic pulse $\br = (x_0,y_0)$. Presume that $V = C_g$ where $C_g$ is the group velocity of a wave of wavenumber $k_g$. This observer would measure waves passing by at a frequency determined by the difference between group and phase velocities for the relevant wavelength. If the phase velocity in this direction for wavenumber $k_g$ is $C$, the observed wave frequency is 
\be\label{eq:freq}
(C-C_g)k_g. 
\ee

The time-dependent wave amplitude $A(t)$ seen by the said observer for an experimentally measured ring wave is calculated as:

\be\label{eq:gv}
A(t) = \hat{\zeta}^{\mathrm{EXP}}(x_0 + (U_0 \pm C_g)t, y_0, t),
\ee
where $U_0$ is the measured surface velocity of the shear current, the $+(-)$ sign for shear-assisted (inhibited) directions in the case of the stagnation region flow and opposite for the curved mesh flow, and $t$ is the time after the initial pulse. The velocity $V = C_g$ of the observer is defined relative to the surface velocity $U_0$. 

The observed wave frequency is found by analyzing $A(t)$ in the frequency domain and finding the peak frequency with a Gaussian fit. The observed wave frequency as a function of observer group velocity in the shear-assisted and shear-inhibited directions for ring waves in the curved mesh and stagnation region flows is shown in figure~\ref{fig:groupvel}. The circles, squares, and triangles represent the frequencies extracted from (\ref{eq:gv}), while the dashed and dotted curves represent theoretical predictions based on the calculated dispersion relation and (\ref{eq:freq}). Also shown in figure~\ref{fig:groupvel}b is the quiescent water case for which the frequencies are the same regardless of the direction of movement of the observer. The data points shown include only wavenumber components where $S(k_g)$ was greater than 50\% of the peak value. In both plots, the frequencies are higher for the shear assisted cases, quantifying the predicted increased (decreased) difference between the phase and group velocities for shear-assisted (inhibited) directions of motion, also evident in the curves in figure~\ref{fig:dr}b. It is noted that the frequency variation is due in part to the variation in $k_g$, which increases (decreases) for the shear-assisted (inhibited) direction. The frequencies for the quiescent water case lie in between the shear-assisted and shear-inhibited cases as expected. Agreement with theoretical predictions is satisfactory in all cases.

\section{Ship wave observations}
\label{sec:ship}

Ship waves were measured in the stagnation flow region with a miniature Series 60 hull of length $L=110$ mm 
and beam $16$ mm
for different values of the Froude number $\Fr = V/\sqrt{gL}$ ($V$: velocity relative to the free surface) and three different angles relative to the stagnation current profile B in figure~\ref{fig:pivih}a. We consider only waves behind the ship, as the capillary waves in front are at least partially obstructed by the ship towing apparatus, and predicted to be very small in wavelength given the values of the surface tension (see e.g. figure~\ref{fig:dr}b), likely smaller than the spatial resolution of the synthetic Schlieren system.

\begin{figure}
  \centering \includegraphics[width=\textwidth]{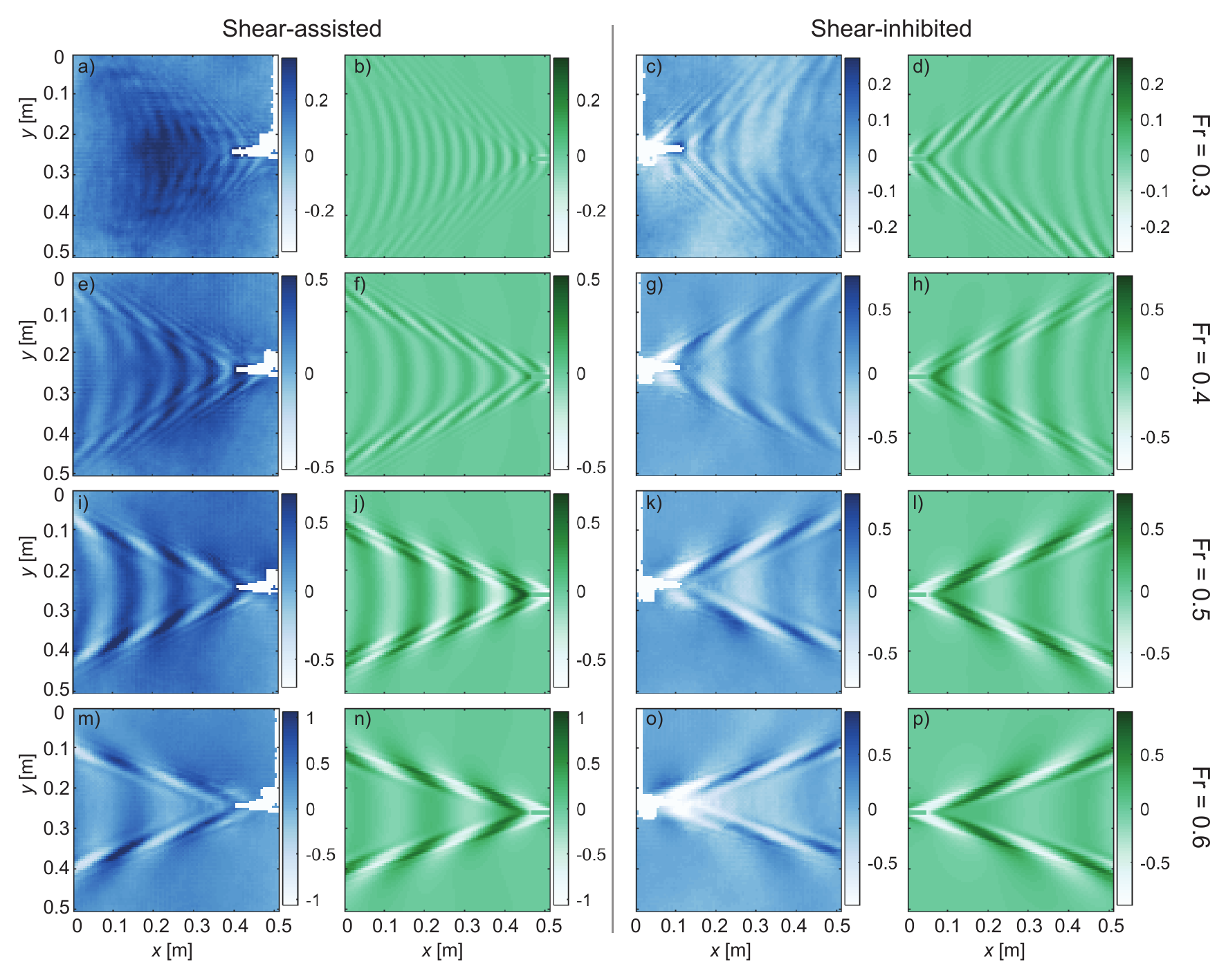}
  \caption{ 
  (colour online) 
    Ship waves in the stagnation flow region at Froude numbers $0.3$, $0.4$, $0.5$, and $0.6$ (rows 1--4 from the top, respectively),
    for ship motion in shear--assisted (Columns 1 and 2 from the left), and
    shear--inhibited (Columns 3 and 4) directions.
    Columns 1 and 3 are lab measurements (blue--tinted); 2 and 4 are theoretical predictions (green--tinted).
    In all panels the surface is at rest in the lab--frame with sub--surface flow towards the right. The colour-bars show the surface elevation in millimeters.
  }
  \label{fig:shipx}
\end{figure}

\begin{figure}
  \centering \includegraphics[width=0.9\textwidth]{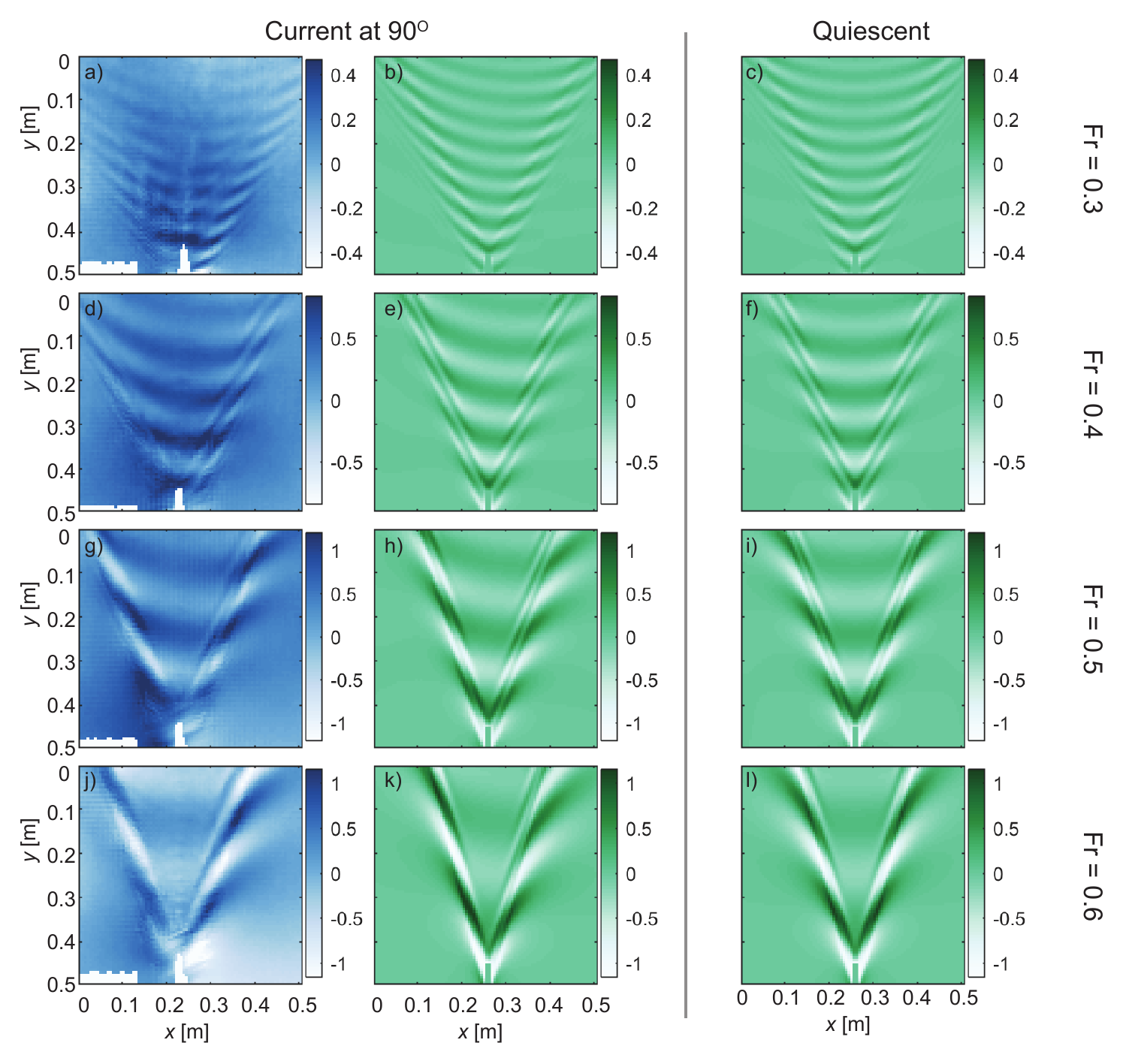}
  \caption{ 
  (colour online) 
    Ship waves in the stagnation flow region for motion in the cross-current direction at Froude numbers $0.3$, $0.4$, $0.5$, and $0.6$ (rows 1--4 from the top, respectively).
    Column 1 are lab measurements (blue--tinted); 2 are theoretical predictions (green--tinted). The theoretical prediction in quiescent waters is shown in Column 3 for comparison. In all panels the surface is at rest in the lab--frame with sub--surface flow towards the right. The colour-bars show the surface elevation in millimeters. 
  }
  \label{fig:shipy}
\end{figure}

The results for ship motion in the shear-assisted and shear-inhibited directions are shown in figure~\ref{fig:shipx}, and for the cross-flow ($90^\circ$) direction in figure~\ref{fig:shipy}.
The measured waves behind the ship are shown in the blue-tinted panels (columns 1 and 3 from the left in figure~\ref{fig:shipx} and column 1 in figure~\ref{fig:shipy}) at four different Froude numbers ---  $\Fr=0.3, 0.4, 0.5$ and $0.6$ --- one per row in the figures as indicated. 
Corresponding theoretical predictions based on the measured velocity profiles and surface tension using the theory described in \S \ref{sec:theory} are shown in the green--tinted panels (columns 2 and 4 of figure~\ref{fig:shipx} and columns 2 and 3 of figure~\ref{fig:shipy}). 
The sub--surface flow is from left to right in all panels in both figures. The coefficient $\Gamma$ in the stagnation region was measured to be $3.7 \pm 0.1 \times 10^{-5}$ $\mathrm{m}^3\mathrm{s}^{-2}$ for the experiments in figure~\ref{fig:shipx} and $2.4 \pm 0.1 \times 10^{-5}$ $\mathrm{m}^3\mathrm{s}^{-2}$ for those in figure~\ref{fig:shipy}.  For the theoretical predictions the ship was modelled as in \S\ref{sec:theory}, equation (\ref{eq:pext}). The theoretical model is fully transient to include the effects of the acceleration phase, using (\ref{zeta}) for a moving pressure distribution. To facilitate a more quantitative comparison between the measured waves and theory, the theoretically predicted surface elevation is normalized such that the RMS of the along-ship gradient component is equal to that measured experimentally. The colour-bars in all panels show the surface elevation in millimeters.

\begin{figure}
  \centering \includegraphics[width=\textwidth]{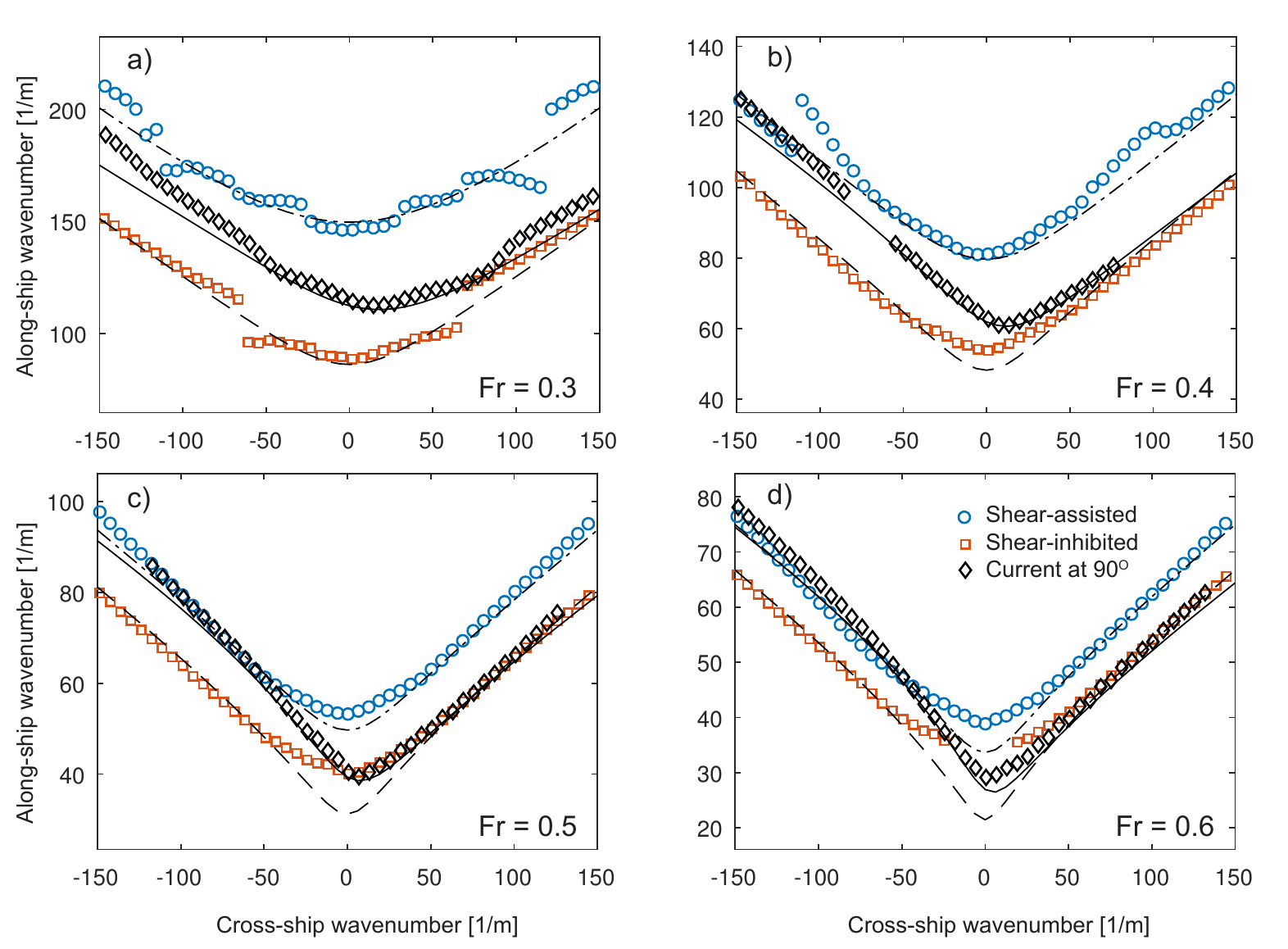}
  \caption{Dispersion curves for stationary ship waves in $\bk$ space extracted from experimental measurements in figure~\ref{fig:shipx} and figure~\ref{fig:shipy} for $Fr = 0.3$, 0.4, 0.5, and 0.6 in a)-d) respectively. The legend applies to a)-d). The dashed and dotted lines show the theoretical predictions.
  }
  \label{fig:statphase}
\end{figure}

A visual comparison shows satisfactory agreement with theoretical predictions. A number of predicted features are clearly visible. Most striking is the asymmetry of the patterns for ship motion at $90^\circ$ in figure~\ref{fig:shipy}, predicted first by \citet{ellingsen14a}. The transverse waves (those directly behind the ship in the central region of the V-shaped wake pattern) make an angle to the direction of ship motion, differing from the quiescent water case as well as the shear-assisted and inhibited directions atop a shear flow. In addition, the width of the diverging wave packets at the boundaries of the wake pattern is wider on the port side of the ship (the right side in figure~\ref{fig:shipy}), most evident for the two highest Froude numbers. Both of the above phenomena are visible in the modeled ship wakes as well (Column 2).
Next, the wavelength of transverse waves is shortened (lengthened) for 
shear-assisted (inhibited) 
motion, and is correctly predicted in all graphs \rev{in figure~\ref{fig:shipx}}. 

The wavelength variation of the transverse waves 
confirms intuition considering the sub-surface shear currents as discussed by \citet{ellingsen14a}. For shear assisted (inhibited) motion, the sub-surface flow is along (against) the direction of the transverse waves, increasing (decreasing) their phase velocity as seen in figure~\ref{fig:dr}b (the same trends occur for shear profile B relevant here). Since $c(\bk)$ is a monotonically decreasing function of $k$ in all directions, the wavelength of a wave having phase velocity equal to the ship speed might be expected to be shorter (longer), confirmed by considering the intersection of the phase velocity curves in figure~\ref{fig:dr}b with a line of constant velocity. 
Differences between the measured and modeled ship wakes in figure~\ref{fig:shipx} and \ref{fig:shipy} are largely attributed to the use of a pressure patch in modeling the ship hull. Firstly, the hypergaussian pressure distribution has a different spatial distribution than the profile of the ship hull, resulting in different weighting of various wavelength components reflected by the $p_{\mathrm{ext}}(\bk)$-term in (\ref{eq:thship}). Secondly, once in motion the surface depression due to the pressure patch will no longer have exactly the same shape as the patch itself.

A more quantitative comparison can be made from the surface elevation transformed to Fourier space in the two spatial dimensions, by extracting and plotting the peak value of the along-ship (longitudinal) wave number $k_\parallel$ as a function of the cross-ship (transverse) wave number $k_\perp$ for each case, shown in figure~\ref{fig:statphase}. The far-field waves away from the ship are given by the poles where the denominator of equation (\ref{eq:thship}) are zero, c.f. \citet{wehausen60}. The spectral peak values of $\bk$ in figure~\ref{fig:statphase} thus correspond to the solution to $c_+(\bk)=0$, the criterion for the waves to be stationary in the ship's system of reference. For a perfectly stationary ship wave pattern, only these $\bk$-values can be present. Extraction is made from a frame average of 
video 
frames where the model ship is in the field of view of the camera, finding the peak along-ship wavenumber as a function of cross-ship wavenumber using a Gaussian fit of the frame-averaged surface elevation magnitude in Fourier space. 

Several predicted qualitative properties observed visually in figures \ref{fig:shipx} and \ref{fig:shipy} are confirmed. The ratio $k_\parallel/k_\perp$ at stationary phase is consistently higher for S.A. than for S.I. ship motion, 
evidence that the dominant wave modes make a steeper angle with the line of motion. In particular, the higher value of $k_\parallel$ at $k_\perp=0$ quantifies the shorter wavelength of transverse waves, which is also easy to see by eye. 
The curve of stationary phase is moreover highly asymmetrical for cross--current motion. The minimal value of the ratio is at a non-zero value of $k_\perp$, indicating that the transverse waves are skewed, as is indeed obvious in figure~\ref{fig:shipy}g and j. 
We also plot the curves expected from theory, based on the numerically evaluated dispersion relation in the stagnation region (dashed and dotted lines). 

The agreement is reasonably good overall, considering the limited 
spatial extent of the measurement area, and hence the limited resolution in $\bk$ space. 
The exception is the smaller wavenumbers in figure~\ref{fig:statphase}c-d 
which we cannot measure accurately for this reason. 
Indications are also present that transient effects play a role for these longest wavelengths which appear in the transverse waves. For shear-inhibited motion in figure \ref{figx} the transverse wavelength of the last, rightmost perid is visibly slightly shorter than those closest to the ship, a remnant of the ring wave generated in the acceleration phase which the ship has not yet completely left behind. This would tend to give a higher value of $k_x$ near $k_y=0$, as observed for S.I.\ especially in figure \ref{fig:statphase}c and d than that expected for a fully stationary ship wave pattern. The effect is greater for the two higher Froude numbers where the average is taken over only two wavelengths of the transverse waves visible in the frame, whereas for lower $Fr$ more wavelengths are present in the region where the ship waves have reached their stationary shape. At $Fr=0.6$ the effect is evident to some extent for all directions of motion for the same reason. 

Deviations for $Fr = 0.3$ and 0.4 cases in figure~\ref{fig:statphase}a-b are likely due to a lower signal-to-noise ratio caused by the smaller wave amplitudes, as well as variation in wave dispersion across the measurement area caused by the surface boundary layer development (see figure~\ref{fig:pivih}b), which increases for smaller $Fr$. Nonetheless, in most cases the deviations between experiment and theory are significantly less than the variations in the curves for different directions of ship motion, confirming the predicted trends. Furthermore, differences in $k_\parallel$ in the shear-assisted versus shear-inhibited directions is significantly greater than the estimated variation in wavenumber over the measurement domain from the surface boundary layer (5\% variation for $Fr = 0.3$ and 2\% for $Fr = 0.6$), ruling out the possibility that the trends are due to the boundary layer.


\section{Concluding remarks} 

We have made laboratory observations of ring waves and ship waves distorted by a sheared current beneath the surface. 
Ring waves and ship waves were created pneumatically on a tailored shear current, and the surface elevation and shear profile were measured using a synthetic Schlieren method and particle image velocimetry, respectively. In a region of strong shear beneath a stagnant surface (the Reynolds ridge phenomenon) very visible shear effects were observed, and effects are also clearly evident for a more weakly sheared, approximately depth-linear current created by a curved mesh at the flow inlet. Ring waves are clearly asymmetrical, and ship waves appear very different in shear--assisted, shear--inhibited and cross--current directions of motion, respectively. The results are compared to recently developed theory for linear surface waves on shear currents of arbitrary shear using the measured velocity profile as input, and the correspondence is qualitatively very satisfactory. 

Also a quantitative comparison of the dispersion curve for stationary ship waves in Fourier space reveals good correspondence considering the level of experimental accuracy.

Quantitative comparisons with theory predictions were performed for both ring waves and ship waves. 
The of measured average phase shift for a ring wave is grossly mispredicted by no-shear theory, but in good agreement with predictions based on the measured shear current. 
Next, the wave frequency seen by an observer moving with the goup velocity is shown to deviate sharply for upstream (shear-inhibited for the stationary flow) vs downstream direction (shear-assisted) for all shear flows in agreement with theory, while it conforms with theory for quiescent water for propagation normal to the shear current, also as expected. 

Peak values of the measured 2-dimensional Fourier spectrum for ship waves are shown to agree well with the predicted criterion of stationary ship waves, with the exception of some cases where results are imperfect due to the limited wave-number resolution, transient effects and/or experimental noise. Once again the standard theory for quiescent water (or, equivalently, depth-uniform current) will cause gross misprediction. 

None of these observed results would be present on a current which is uniform in depth.

Our results not only confirm theoretically predicted effects for these canonical surface wave patterns, but also lend strong 
empirical support to the reality of a range of profound real-life consequences. In particular, the same theory was recently used by \cite{li17} to predict wave resistance on ships in the Columbia river delta, finding a difference of a factor 3 in wave resistance for upstream vs downstream motion at the same velocity relative to the water surface, typical of smaller ships trafficking these waters. Also, highy significant lateral wave radiation forces as well as transient effects during manoeuvring were predicted. Similar conclusions for the case of hurricane waves were drawn long ago by \cite{dalrymple73}. The necessity to include shear effects in calculation of wave loads and forces on vessels and installations is re-emphasized. 
There exists at present no practical and affordable 
design tool for predicting the three-way behaviour of physical systems involving both waves, moving bodies and shear currents: most calculational tools (e.g.\ the WAMIT toolkit \cite{lee95}) are developed for potential flow and are inapplicable in the presence of shear \citep{ellingsen16}, and computational fluid mechanics solving the full Euler or Navier-Stokes equations, as well as laboratory experiments with shear flow are typically too costly and time--consuming for design purposes involving numerous parameters. An exception we are aware of is the CFDShip-Iowa code \citep{paterson03} which, being a Navier-Stokes solver allows rotational flow; no calculations appear to have been made for ship waves and loads on a shear current to date, however.
The development of efficient predictive tools for wave-body-shear current behaviour should therefore be a priority in the near future. 

\subsection*{Acknowledgements}
S\AA E is funded by the Research Council of Norway (FRINATEK), grant no.\ 249740. 
Experiments were performed with the invaluable assistance of Anna \AA dn\o y and Stefan Weichert.

\bibliographystyle{jfm}

\end{document}